\documentclass[aps,prl,reprint]{revtex4-1}
\usepackage{blindtext}

\pdfoutput=1
\newcommand{\beq}{\begin{equation}}
\newcommand{\eeq}{\end{equation}}
\usepackage{graphicx}
\usepackage{natbib}

\usepackage{hyperref}

\usepackage{amsmath}

\usepackage{multirow} 
\usepackage{tabularx}
\usepackage{xcolor}

\begin{document}
\title{A unified model of inflation and dark energy based on the holographic spacetime foam}
\author{Daniel Jim\'enez-Aguilar}
\email{daniel.jimenez@ehu.eus}
\affiliation{Department of Physics, UPV/EHU, 48080, Bilbao, Spain}

\begin{abstract}
I present a model of inflation and dark energy in which the inflaton potential is constructed by imposing that a scalar field representing the classical energy of the spacetime foam inside the Hubble horizon is an exact solution to the cosmological equations. The resulting potential has the right properties to describe both the early and late expansion epochs of the universe in a unified picture.
\end{abstract}

\maketitle

\section{1. Introduction}
Recent cosmological observations indicate that the universe is currently experiencing an accelerated expansion \cite{Riess, Perlmutter, Spergel, Tegmark}. This fact can be attributed to the existence of some form of energy (dubbed dark energy) with negative pressure. Our unawareness of the nature of dark energy is certainly at the root of one of the greatest unknowns in theoretical physics: the cosmological constant problem \cite{Weinberg}. On the other hand, it is widely accepted today that the very early universe underwent a period of quasi-exponential expansion called inflation, which can address many of the problems of the standard Big Bang cosmology \cite{Guth, Starobinsky, Kazanas, Sato, Linde, AlbrechtAndSteinhardt}. \\\\
Both expansion epochs may be traced to a common cause by interpreting dark energy as a dynamical scalar field (the inflaton) that slowly rolls down its potential to drive inflation in the early universe and finally resembles a cosmological constant at the present time, at a much lower energy scale. Depending on the shape of the potential, the field will end up oscillating about a minimum (as in the model presented in this paper) or rolling down an infinite tail as quintessence. The latter scenario is called quintessential inflation \cite{Peebles}. \\\\
Another chance of uncovering the nature of dark energy may be found in the structure of spacetime itself. Due to its quantum nature, spacetime is foamy on scales of the order of the Planck length \cite{Gravitation}. The quantum fluctuations of the metric are responsible for a perpetual change in the geometry of spacetime, and consequently, any measurement of space and time intervals acquires some uncertainty. In a region of spatial extent $L$, this uncertainty is given by the K\'arolyh\'azy relation \cite{Karolyhazy}, which was derived independently by other authors \cite{Sasakura, Jack1994, Jack2006, Jack2007}:
\begin{equation}
\delta L\sim L_{p}^{2/3}L^{1/3},
\label{eq:karolyhazy}
\end{equation}
where $L_{p}$ is the Planck length. Note that $\delta L\sim L$ precisely at the Planck scale. This picture of the small scale structure of spacetime is called spacetime foam. An appealing idea that has been proposed is that the energy density associated to the spacetime foam is the one that drives both the early and late expansions of the universe \cite{Jack2021}.\\\\
The K\'arolyh\'azy uncertainty relation (\ref{eq:karolyhazy}) is closely related to the holographic principle \cite{Hooft, Susskind}, as it establishes a connection between the ultraviolet ($\delta L$) and infrarred ($L$) cut-off scales of the system. Indeed, expression (\ref{eq:karolyhazy}) is also known as the holographic spacetime foam model, as it suggests that the number of degrees of freedom or bits of information in that region is proportional to its surface area: $\left(L/\delta L\right)^{3}\propto L^{2}$. The ultraviolet cut-off scale is related to the energy density of the vacuum, and one is led to a holographic dark energy of the form
\begin{equation}
\rho\sim\frac{1}{L_{p}^{2}L^{2}}\,.
\label{eq:HDE}
\end{equation}
The usual approach in holographic dark energy models is to propose an ansatz for the infrarred cut-off $L$ (some natural choices are the Hubble radius \cite{Hsu, ZimdahlandPavon}, the event horizon \cite{Li}, the age of the universe \cite{Cai} and the Ricci length \cite{DuranandPavon}) and combine equation (\ref{eq:HDE}) with the Friedmann equation in order to extract the Hubble rate as a function of time. Then, one can see whether this particular choice leads to an accelerated expansion, and also constrain the model with the observational data. While these models have been intensively applied to the late universe (see \cite{delCampo} for a review), only a few authors have extended these ideas to the early universe (for instance, see \cite{HolographicInflation}). \\\\
Finally, and more on the line of thought of this work, there have been several studies aimed at establishing a connection between holographic dark energy and scalar field models \cite{Zhang, Karami, Cui, Sheykhi2010}, although the explicit reconstruction of the scalar field potential has proved to be challenging (however, see \cite{Sheykhi2011, Granda}). \\\\
In this paper, I present a unified model of inflation and dark energy in which the scalar potential is reconstructed by making a specific ansatz for the field in the cosmological equations. This ansatz will be rooted at the holographic model of spacetime foam.\\\\
This letter is structured as follows: in section 2, I expose the main idea behind the construction of the inflaton potential. This idea is applied in section 3 under the assumption that the scalar field is real. In this case, the potential obtained is not satisfactory. The calculation is redone in section 4 for a complex scalar field, obtaining a family of acceptable potentials. In section 5, I pick up the potential that has direct connection with the spacetime foam and I discuss some of its properties. Finally, some concluding remarks are made in section 6.\\\\
Natural units ($c=\hbar\equiv1$) are used throughout the letter. In particular, this implies that Newton's gravitational constant is given by $G=L_{p}^2=M_{p}^{-2}$, where $M_{p}$ is the Planck mass.

\section{2. The main idea}
Regardless of the initial field configuration in the patch of the universe that is going to inflate, the field evolves under the influence of a potential $V\left(\phi\right)$. The Friedmann and Klein-Gordon equations in a spatially flat Friedmann-Lema\^itre-Robertson-Walker (FLRW) universe establish a correspondence between the potential and a homogeneous function of time $\phi\left(t\right)$. Therefore, one can obtain the scalar potential by imposing that a particular $\phi_{s}\left(t\right)$ is an exact solution to these equations (the subscript $s$ stands for \emph{solution}). For instance, $V\left(\phi\right)$ could be determined by imposing that $\phi_{s}\left(t\right)\propto H\left(t\right)$ is a solution to the equations (here and henceforth, $H$ denotes the Hubble rate). Whatever our ansatz is, $\phi_{s}\left(t\right)$ is not the actual inflaton field $\phi\left(t,\vec{x}\right)$ in the universe, but just a tool to determine the potential. Indeed, a natural initial state for the field is arbitrary and inhomogeneous. Since $\phi\left(t=0,\vec{x}\right)\neq\phi_{s}\left(t=0\right)$ and $\dot{\phi}\left(t,\vec{x}\right)\neq\dot{\phi}_{s}\left(t=0\right)$, the inflaton field would simply end up settling in the vacuum of the potential, possibly resembling the cosmological constant.\\\\
Here we will obtain a family of scalar field potentials by considering the simplest ansatz with dimensions of energy that one can make out of $M_{p}$ and $H$:
\begin{equation}
\phi_{s}\left(H\right)=AM_{p}^{\gamma}H^{1-\gamma},
\label{eq:phi solution}
\end{equation}
where $A$ is a dimensionless constant and $\gamma$ is a generic exponent that will label the different elements of the family of solutions. Some of them will not only have the right shape to allow for inflation in the early universe and cosmological constant at the present time, but also a connection with the spacetime foam, as we point out now.\\\\
Note that ansatz (\ref{eq:phi solution}) for $\gamma=2/3$ is essentially the inverse of the K\'arolyh\'azy uncertainty relation (\ref{eq:karolyhazy}) evaluated at the Hubble scale. One can show that $1/\delta L$ is proportional to the energy of the spacetime foam inside the Hubble horizon. The classical energy density associated to the metric fluctuations is given by \cite{Maziashvili2006}
\begin{equation}
\rho_{foam}\sim\frac{1}{L_{p}^{2/3}L^{10/3}},
\label{eq:classical foam energy density}
\end{equation}
and although this expression was derived in Minkowski spacetime, we will assume that the powers of $L_{p}$ and $L$ remain unchanged in a spatially flat FLRW universe. If the scale $L$ in equation (\ref{eq:classical foam energy density}) is chosen to be the Hubble radius, the classical energy of the spacetime foam in that volume is
\begin{equation}
E_{foam}\sim M_{p}^{2/3}H^{1/3}.
\label{eq:classical foam energy}
\end{equation}
This expression for the energy can also be derived from the Margolus-Levitin theorem in quantum computation \cite{Margolus}. The dynamical evolution of any physical system can be thought of as a succession of orthogonal quantum states, and each step between consecutive states can be understood as an elementary operation. It can be easily shown that each of these steps takes at least time $\delta t\sim E^{-1}$, where $E$ is the average energy of the system. Therefore, $E\sim\delta t^{-1}$, which yields expression (\ref{eq:classical foam energy}) if we impose that $\delta t$ corresponds to the uncertainty given by the K\'arolyh\'azy relation (\ref{eq:karolyhazy}) evaluated at the Hubble scale.

\section{3. Real scalar field}
\begin{figure}[h!]
\includegraphics[width=0.48\textwidth]{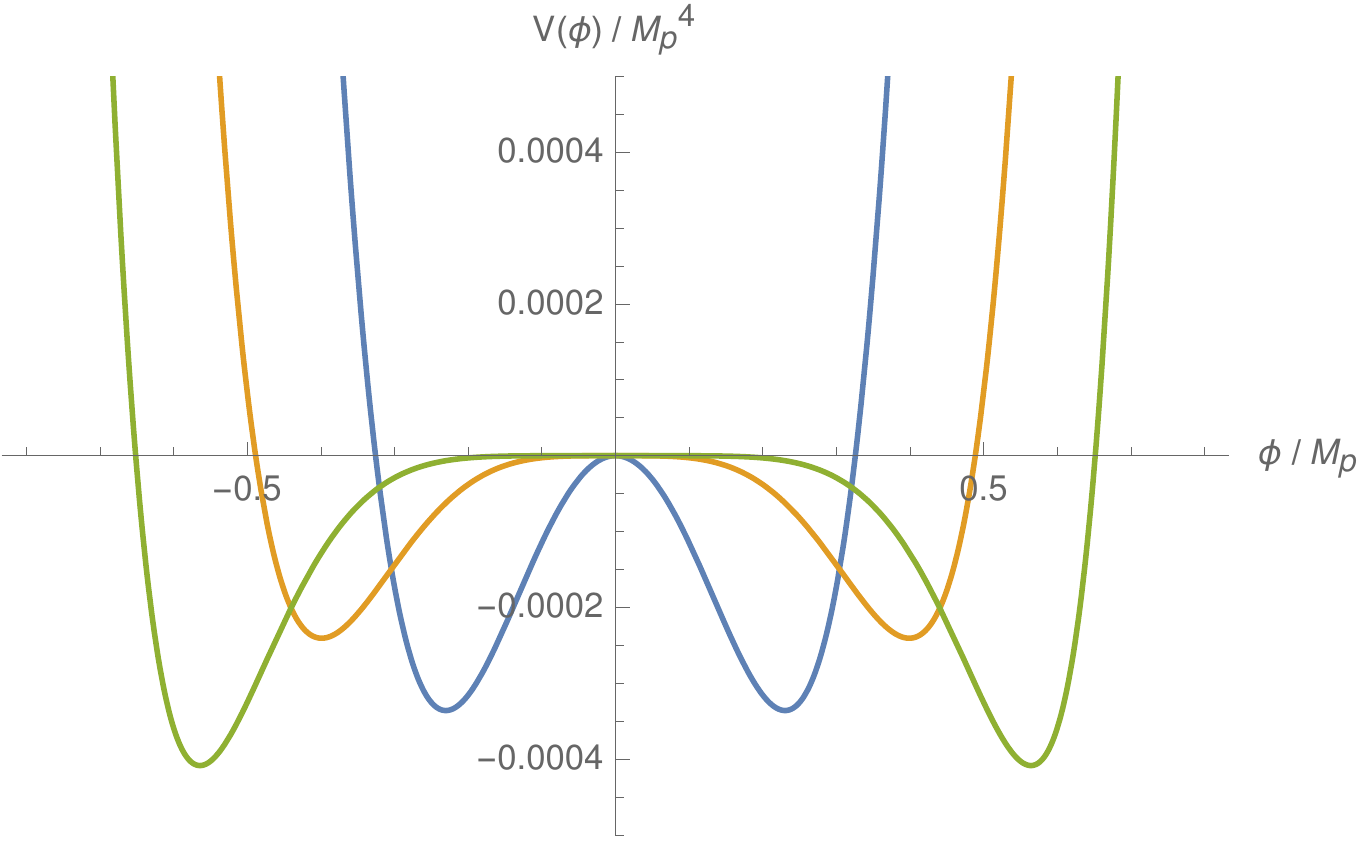}
\caption{Potential (\ref{eq:V as a function of phi}) for $\gamma=1/2$ (blue), $\gamma=2/3$ (orange) and $\gamma=3/4$ (green). In this plot, we have set $A=1$.}
\label{fig:even potential}
\end{figure}

\begin{figure}[h!]
\includegraphics[width=0.48\textwidth]{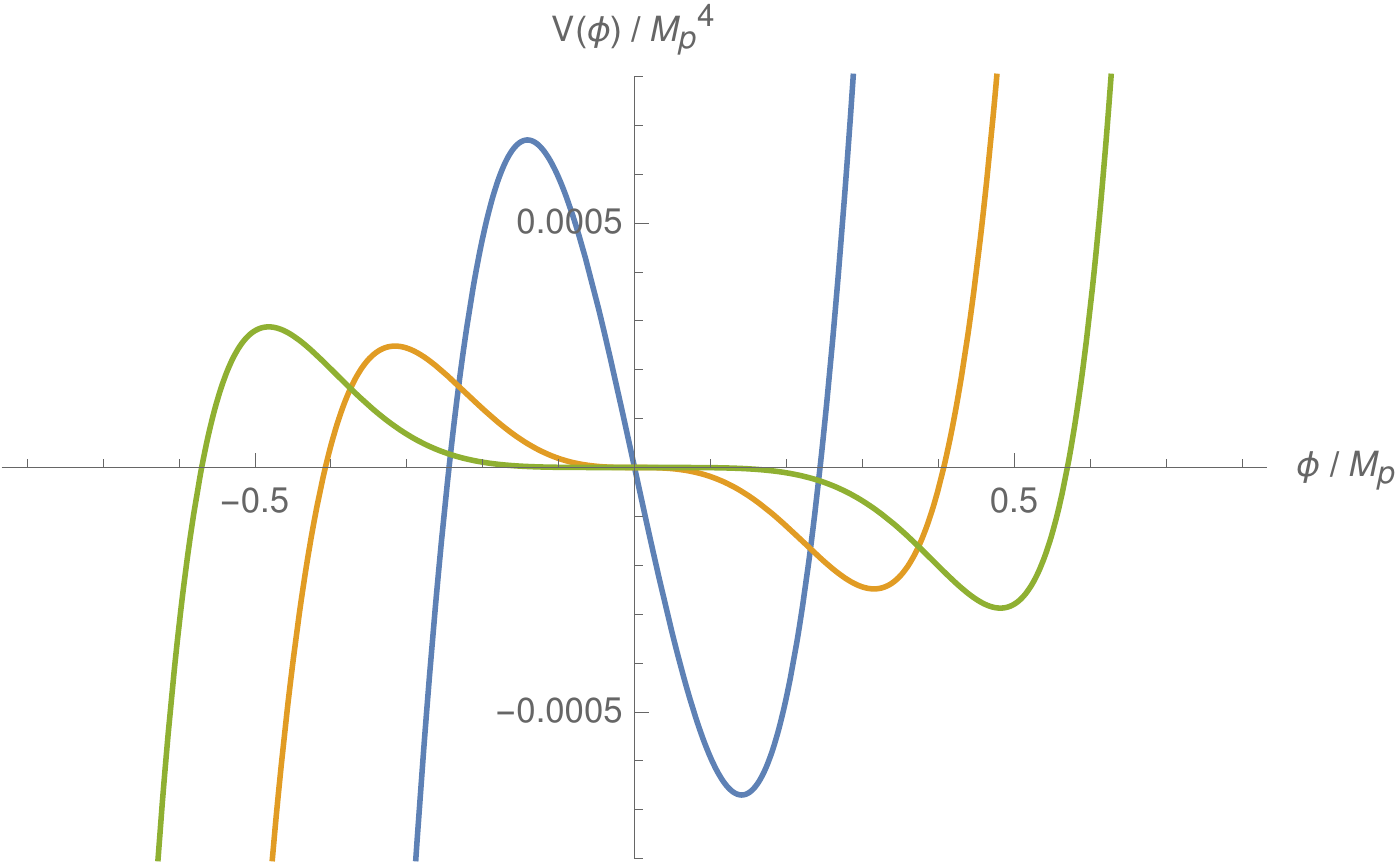}
\caption{Potential (\ref{eq:V as a function of phi}) for $\gamma=1/3$ (blue), $\gamma=3/5$ (orange) and $\gamma=5/7$ (green). In this plot, we have set $A=1$.}
\label{fig:odd potential}
\end{figure}

The simplest possibility is to consider a real scalar field. The first Friedmann equation and the Klein-Gordon equation read
\begin{equation}
H^{2}=\frac{8\pi}{3M_{p}^{2}}\left[\frac{\dot{\phi}^{2}}{2}+V\left(\phi\right)\right],
\label{eq:friedmann}
\end{equation}
\begin{equation}
\ddot{\phi}+3H\dot{\phi}+\frac{dV}{d\phi}=0.
\label{eq:KG}
\end{equation}
Taking the time derivative of (\ref{eq:friedmann}) and combining it with (\ref{eq:KG}) yields
\begin{equation}
\dot{H}=-\frac{4\pi}{M_{p}^{2}}\dot{\phi}^{2}.
\label{eq:KG like}
\end{equation} 
Since $\phi=\phi\left(t\right)$ and $H=H\left(t\right)$, one can take $H=H\left(\phi\right)$. Taking this into account, and using equations (\ref{eq:friedmann}) and (\ref{eq:KG like}), one can express $V\left(\phi\right)$ as
\begin{equation}
V\left(\phi\right)=\frac{M_{p}^{2}}{8\pi}\left[3H^{2}\left(\phi\right)-\frac{M_{p}^{2}}{4\pi}\left(\frac{dH}{d\phi}\right)^{2}\right]\,.
\label{eq:potential}
\end{equation} 
Now we determine $V\left(\phi\right)$ by imposing that $\phi=\phi_{s}\left(H\right)$, given by equation (\ref{eq:phi solution}). Inverting to get $H\left(\phi\right)$ and substituting in (\ref{eq:potential}) yields
\begin{equation}
V\left(\phi\right)=\frac{M_{p}^{\frac{2\left(1-2\gamma\right)}{1-\gamma}}}{8\pi A^{\frac{2}{1-\gamma}}}\left[3\phi^{\frac{2}{1-\gamma}}-\frac{M_{p}^{2}}{4\pi\left(1-\gamma\right)^{2}}\phi^{\frac{2\gamma}{1-\gamma}}\right]\,.
\label{eq:V as a function of phi}
\end{equation} 
We further assume that $V\left(\phi\right)$ takes finite values for finite $\phi$. Together with the fact that the potential is real, this implies that the exponent $\frac{2\gamma}{1-\gamma}\equiv n$ is a natural number (if this is the case, the first exponent is also a natural number: $\frac{2}{1-\gamma}=n+2$). This means that $\gamma$ is quantized: $\gamma=\frac{n}{2+n}$. As shown in figures \ref{fig:even potential} and \ref{fig:odd potential}, the potential is either symmetric (if $n$ is even) or antisymmetric (if $n$ is odd). For $\gamma=0$, the potential is a parabola with its minimum at $\phi=0$, at a negative energy density.\\

In the case of the symmetric double-well potentials, note that the flatness around $\phi=0$ is controlled by $\gamma$. Note also that the minima are placed at negative values of the potential energy density. Therefore, none of them can be the actual potential for the inflaton. However, if one considers a complex scalar field, the potential contains an extra term related to the phase. This extra term can act as an uplifting term, so one can obtain vacua with positive cosmological constant in that case.

\section{4. Complex scalar field}
If we consider a complex field $\phi=\frac{\varphi}{\sqrt{2}}e^{i\frac{\beta}{\varphi_{v}}}$, where $\beta$ is the angular part of the field and $\varphi_{v}$ is the vacuum expectation value of the radial part, equations (\ref{eq:friedmann}) and (\ref{eq:KG like}) read
\begin{equation}
H^{2}=\frac{8\pi}{3M_{p}^{2}}\left[\frac{\dot{\varphi}^{2}}{2}+\frac{\varphi^{2}\dot{\beta}^{2}}{2\varphi_{v}^{2}}+V\left(\varphi\right)\right],
\label{eq:friedmann complex}
\end{equation}
\begin{equation}
\dot{H}=-\frac{4\pi}{M_{p}^{2}}\left(\dot{\varphi}^{2}+\frac{\varphi^{2}\dot{\beta}^{2}}{\varphi_{v}^{2}}\right).
\label{eq:KG like complex}
\end{equation}
Again, we assume that $\varphi$ is a function of $H$ and hence $H=H\left(\varphi\right)$. Combining equations (\ref{eq:friedmann complex}) and (\ref{eq:KG like complex}), one gets
\begin{equation}
V\left(\varphi\right)=\frac{M_{p}^{2}}{8\pi}\left[3H^{2}\left(\varphi\right)-\frac{M_{p}^{2}}{4\pi}\left(\frac{dH}{d\varphi}\right)^{2}-\frac{\varphi^{2}\dot{\beta^{2}}}{\varphi_{v}^{2}\dot{\varphi}}\frac{dH}{d\varphi}\right]\,.
\label{eq:potential complex}
\end{equation} 
This potential has the same form as (\ref{eq:potential}), except for the presence of a new term that contains the time derivative of the angular part of the field. Note that this extra term can be written as $-\frac{\varphi^{2}\dot{\beta^{2}}}{\varphi_{v}^{2}\dot{\varphi}^{2}}\dot{H}$, and this is always positive as $\dot{H}<0$ (see equation (\ref{eq:KG like complex})). If the phase $\beta$ is chosen in such a way that 
\begin{equation}
\dot{\beta}^{2}\propto\frac{dH^{-1}}{dt}\,,
\label{eq:ansatz for beta}
\end{equation}
that is, proportional to the slow-roll parameter $\epsilon$, the extra term is a constant that uplifts the potential found in the previous section. Note that this choice of $\beta$ is made at the same level as ansatz (\ref{eq:phi solution}) for $\varphi$. The phase determined by equation (\ref{eq:ansatz for beta}), as well as any other $\beta$ that one could have come up with, is an exact solution of the equations of motion provided that the potential energy density is given by (\ref{eq:potential complex}). Other choices of $\beta$ could also do the job of uplifting the potential by locally deforming it, but in these cases the deformation would not be uniform in $\varphi$. This uniformity condition is obviously not necessary to obtain a potential with the right properties to account for the expansion history of the universe, but let us focus on this possibility for simplicity. Making use of ansatz (\ref{eq:ansatz for beta}), the extra term in the potential is a positive constant with dimensions of energy squared: $-\frac{\varphi^{2}\dot{\beta^{2}}}{\varphi_{v}^{2}\dot{\varphi}}\frac{dH}{d\varphi}\equiv BM_{p}^{2}$, where $B>0$ is a dimensionless constant. \\\\
Now we impose that $\varphi=\varphi_{s}\left(H\right)=AM_{p}^{\gamma}H^{1-\gamma}$. Finding $H\left(\varphi\right)$ and substituting in (\ref{eq:potential complex}), one gets the following potential: 
\begin{equation}
V\left(\varphi\right)=\frac{M_{p}^{\frac{2\left(1-2\gamma\right)}{1-\gamma}}}{8\pi A^{\frac{2}{1-\gamma}}}\left[3\varphi^{\frac{2}{1-\gamma}}-\frac{M_{p}^{2}}{4\pi\left(1-\gamma\right)^{2}}\varphi^{\frac{2\gamma}{1-\gamma}}\right]+\frac{BM_{p}^{4}}{8\pi}\,.
\label{eq:V as a function of phi complex}
\end{equation}
For the values of $\gamma$ corresponding to even $n=\frac{2\gamma}{1-\gamma}$, this is a Mexican-hat potential with the circle of minima at $\varphi=\varphi_{v}=M_{p}\sqrt{\frac{\gamma}{12\pi\left(1-\gamma\right)^{2}}}$. The vacuum energy density can be shown to be
\begin{equation}
V\left(\varphi_{v}\right)=-\frac{M_{p}^{2}m^{2}}{48\pi}+\frac{BM_{p}^{4}}{8\pi}\,,
\label{eq:vacuum energy}
\end{equation}
where $m=f\left(\gamma\right)M_{p}$ is the mass of small radial fluctuations about the vacuum, i.e., the mass of the inflaton field. The function $f\left(\gamma\right)$ is given by 
\begin{equation}
f\left(\gamma\right)=\left[\frac{\gamma^{\gamma}}{8^{1-\gamma}\times12^{2\gamma-1}\pi\left(1-\gamma\right)^{1+\gamma}A^{2}}\right]^{\frac{1}{2\left(1-\gamma\right)}}.
\label{eq:function of gamma}
\end{equation}
We want the vacuum energy density (\ref{eq:vacuum energy}) to be equal to the dark energy density today, namely, $\rho_{\Lambda}^{0}=\Omega_{\Lambda}^{0}\rho_{c}^{0}=\frac{3\Omega_{\Lambda}^{0}}{8\pi}H_{0}^{2}M_{p}^{2}$. This implies that 
\begin{equation}
B=3\Omega_{\Lambda}^{0}\left(\frac{H_{0}}{M_{p}}\right)^{2}+\frac{1}{6}\left(\frac{m}{M_{p}}\right)^{2}\approx\frac{1}{6}\left(\frac{m}{M_{p}}\right)^{2}\,.
\label{eq:B}
\end{equation}
The first term is negligible, but its presence makes this choice of $B$ extremely fine-tuned. Evidently, since the model does not provide a fundamental explanation for it, the cosmological constant problem is not resolved. However, once this term is neglected, the approximate value of $B$ is arguably natural: condition (\ref{eq:B}) implies that the Hubble rate during inflation ($H_{I}$) is given by 
\begin{equation}
H_{I}\approx\sqrt{\frac{8\pi}{3M_{p}^{2}}V\left(0\right)}=M_{p}\sqrt{\frac{B}{3}}\approx\frac{m}{3\sqrt{2}}\,.
\label{eq:Hubble rate inflation}
\end{equation}
This means that the Hubble radius during inflation is of the same order of magnitude as the Compton wavelength of the inflaton, which is the only fundamental length scale in the problem apart from the Planck length. Therefore, although the model requires fine-tuning to explain the exact value of the cosmological constant, it may not need it to account merely for its smallness.

\section{5. The case $\gamma=2/3$}
\begin{figure}[h!]
\includegraphics[width=0.48\textwidth]{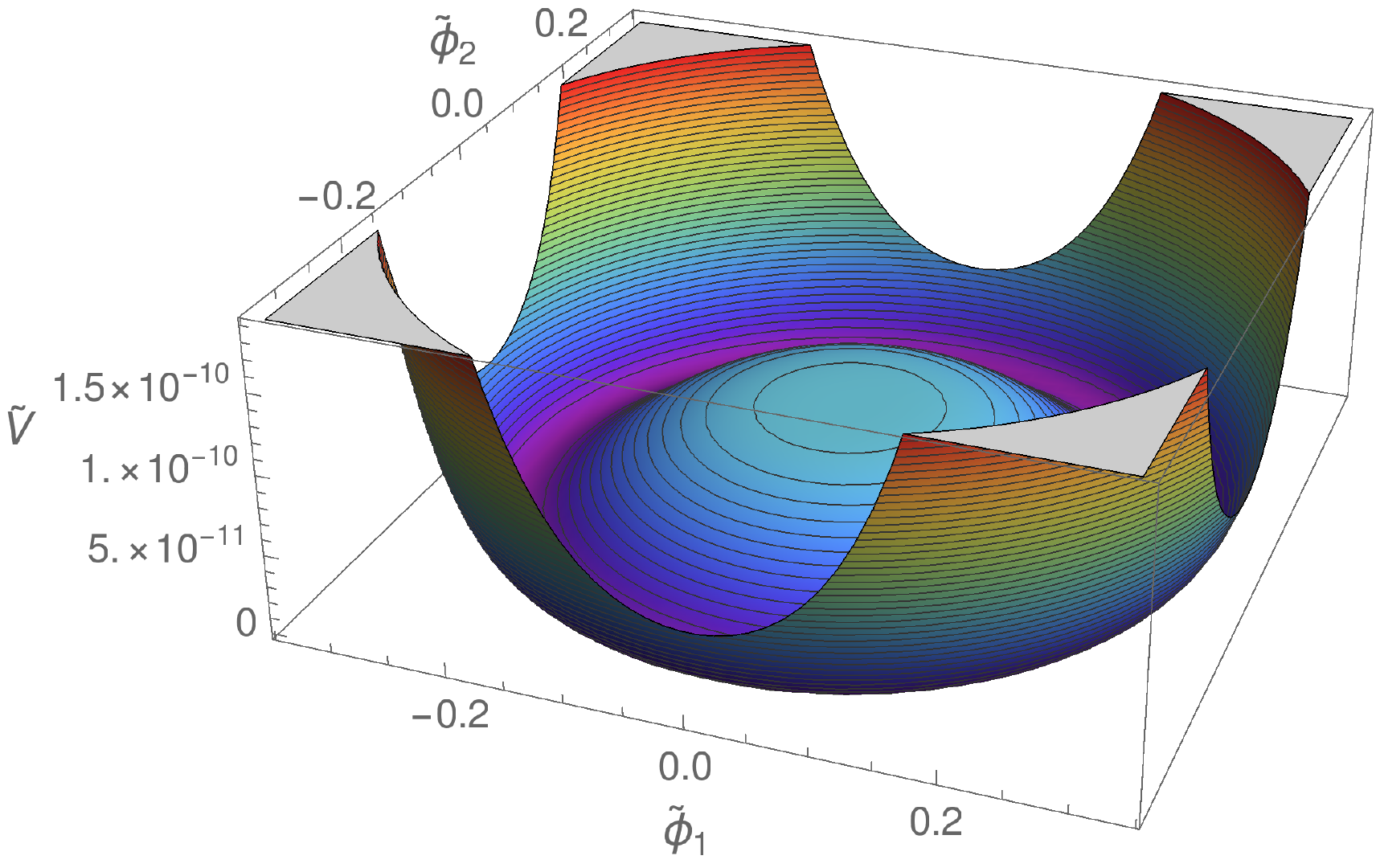}
\caption{Potential (\ref{eq:final potential}) for $m=10^{-4}M_{p}$. $\tilde{\phi}_{1}$ and $\tilde{\phi}_{2}$ correspond to the real and imaginary parts of $\tilde{\phi}=\phi/M_{p}$, respectively.}
\label{fig:plot final potential}
\end{figure}

As previously stated, the ansatz for the radial part of the scalar field can be interpreted as the energy of the spacetime foam inside the Hubble horizon if $\gamma=2/3$. Defining $\tilde{\varphi}\equiv\varphi/M_{p}$ and $\tilde{V}\equiv V/M_{p}^{4}$, and making use of (\ref{eq:function of gamma}) and (\ref{eq:B}), the potential (\ref{eq:V as a function of phi complex}) for $\gamma=2/3$ can be written as 
\begin{equation}
\tilde{V}\left(\tilde{\varphi}\right)\approx\left(\frac{m}{M_{p}}\right)^{2}\left[\frac{\pi^{2}}{3}\left(\tilde{\varphi}^{6}-\frac{3}{4\pi}\tilde{\varphi}^{4}\right)+\frac{1}{48\pi}\right]\,.
\label{eq:final potential}
\end{equation}
With this potential, plotted in figure \ref{fig:plot final potential}, ansatz (\ref{eq:phi solution}) with $\gamma=2/3$ for $\varphi$ and ansatz (\ref{eq:ansatz for beta}) for $\beta$ are exact solutions of the equations of motion. This means that, until radiation and matter become important and influence the dynamics of the scalar field, $\varphi$ and $\beta$ would evolve as dictated by their corresponding ansätze as long as the initial conditions for both of them are exactly given by these expressions. However, we find it more natural to consider the possibility that the initial state is arbitrary and inhomogeneous, and that the scalar field is in different regions of the potential in different Hubble patches. In this case, $\varphi$ and $\beta$ will definitely not follow (\ref{eq:phi solution}) and (\ref{eq:ansatz for beta}), as the evolution of the scalar field will proceed essentially as in a cosmological phase transition. As we will shortly see, a patch of the universe where the radial part of the scalar field starts out sufficiently close to $\varphi\left(t=0,\vec{x}\right)=0$ in potential (\ref{eq:final potential}) will undergo inflation. After that, the field will reach the circle of minima at $\varphi=\varphi_{v}$ and will start oscillating as dictated by the Friedmann and Klein-Gordon equations. Production of the standard model particles will then take place and the universe will be subsequently dominated by radiation and matter. In the long term, the residual vacuum energy (\ref{eq:vacuum energy}) of the field will start dominating as a cosmological constant.    \\

Let us note that, even if the initial conditions were those determined by the ansätze (\ref{eq:phi solution}) and (\ref{eq:ansatz for beta}), quantum fluctuations on top of that state may spoil the subsequent expected evolution and yield a picture which would be similar to the one presented above. \\\\
The properties of the $\gamma=2/3$ potential (\ref{eq:final potential}) are studied quantitatively in the following subsections.

\subsection{5.1. Mass of the inflaton}
\begin{figure}[h!]
\includegraphics[width=0.48\textwidth]{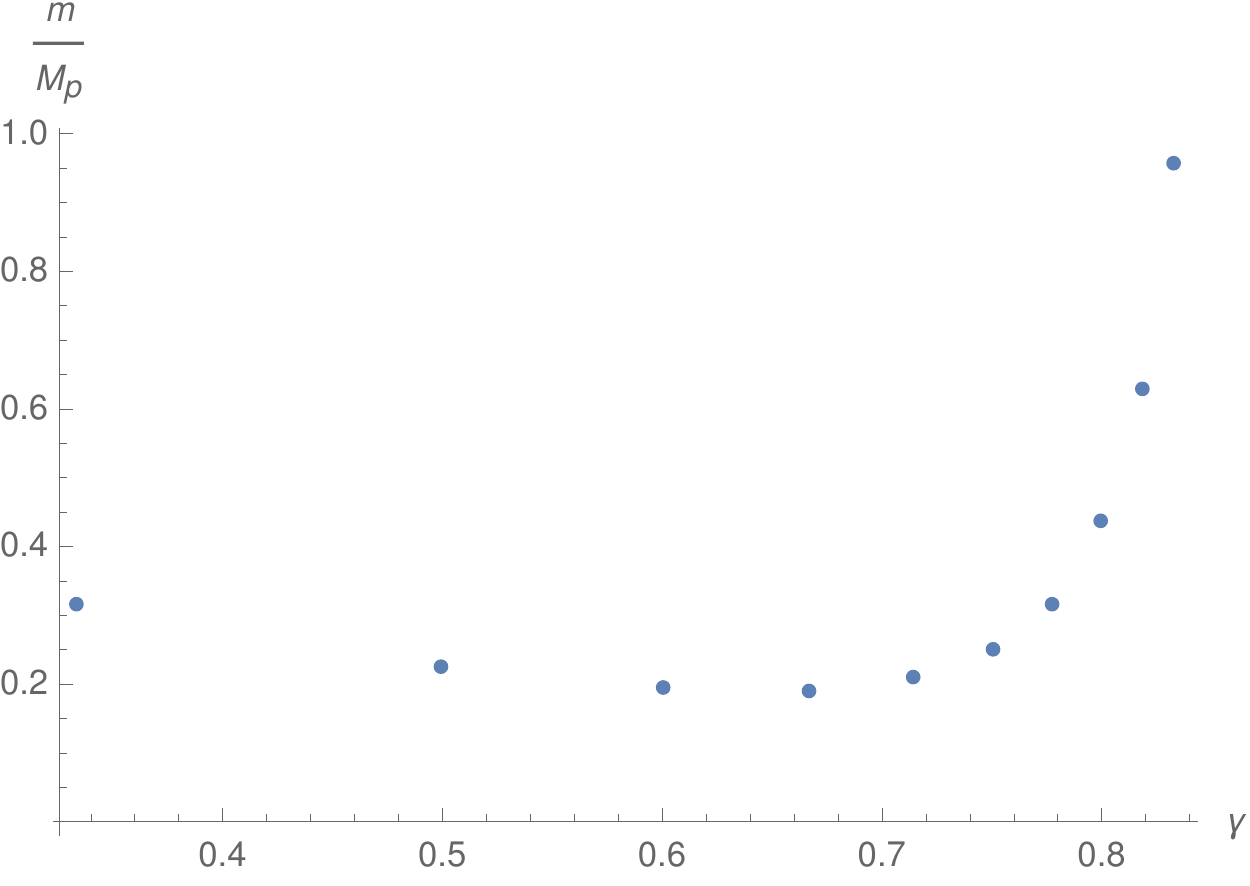}
\caption{Inflaton mass (rescaled by the Planck mass) as a function of $\gamma$ (equation (\ref{eq:function of gamma})). In this plot, we have set $A=1$.}
\label{fig:mass of the inflaton}
\end{figure}

As it is explicit in equation (\ref{eq:function of gamma}), the mass of the inflaton depends on $\gamma$. Since $\gamma$ is quantized, the ratio $m/M_{p}$ can only take discrete values. Interestingly, out of all the possibilities, the minimum inflaton mass is obtained precisely for $\gamma=2/3$ (see figure \ref{fig:mass of the inflaton}).\\

Our model gives a prediction for the inflaton mass (equation (\ref{eq:function of gamma})), but the parameter $A$ is uncertain as we do not know the exact proportionality factor for the energy of the spacetime foam (equation (\ref{eq:classical foam energy})). Considering a conservative choice like $A<20$, we get $m>10^{-5}M_{p}$ as an order of magnitude estimate for the minimum inflaton mass. On the other hand, according to Planck 2018 results \cite{Planck2018}, the Hubble rate during inflation has to verify $H_{I}<2.5\times10^{-5}M_{p}$. Combining this observational constraint with equation (\ref{eq:Hubble rate inflation}), we obtain $m<1.1\times10^{-4}M_{p}$.

\subsection{5.2. Number of e-folds}
Given the inflaton potential (equation (\ref{eq:final potential})), and assuming slow-roll conditions, one can calculate the number of e-folds of inflation as
\begin{equation}
N\left(\varphi\right)=\frac{8\pi}{M_{p}^{2}}\int_{\varphi}^{\varphi_{i}}\frac{V\left(\hat{\varphi}\right)}{V'\left(\hat{\varphi}\right)}\,d\hat{\varphi},
\label{eq:number of e-folds}
\end{equation}
where $\varphi_{i}\sim H_{I}<2.5\times10^{-5}M_{p}$ is the initial value of the inflaton field. Its value at the end of inflation, $\varphi_{end}\approx0.15M_{p}$, is chosen in such a way that the greatest of the two slow-roll parameters $\epsilon$ and $|\eta|$ is approximately $1$: 
\begin{equation}
\epsilon\left(\varphi\right)=\frac{M_{p}^{2}}{16\pi}\left[\frac{V'\left(\varphi\right)}{V\left(\varphi\right)}\right]^{2}\,,
\label{eq:slow roll epsilon}
\end{equation}
\begin{equation}
|\eta\left(\varphi\right)|=\left|\frac{M_{p}^{2}}{8\pi}\frac{V''\left(\varphi\right)}{V\left(\varphi\right)}\right|\,.
\label{eq:slow roll eta}
\end{equation}
The result for $\gamma=2/3$ is $N\left(\varphi_{end}\right)\sim10^{8}$. Therefore, the $\gamma=2/3$ potential allows for more than enough expansion.\\\\
The number of e-folds has been also computed for the cases $\gamma=1/2$ and $\gamma=3/4$. The results are $N\left(\varphi_{end}\right)\sim1$ and $N\left(\varphi_{end}\right)\sim10^{16}$, respectively.\\\\
Note that the previous results are independent of the choice of parameters in the model: after adjusting the potential uplift with (\ref{eq:B}), they only enter via $m$ in (\ref{eq:final potential}), which cancels out exactly in the ratios appearing in (\ref{eq:number of e-folds}), (\ref{eq:slow roll epsilon}) and (\ref{eq:slow roll eta}). 

\subsection{5.3. Tensor-to-scalar ratio and scalar tilt}
Let $N_{*}$ be the number of e-folds between horizon crossing for observable wavelengths and the end of inflation: $N_{*}=N\left(\varphi_{end}\right)-N\left(\varphi_{hc}\right)$. Here, the subscript $hc$ stands for \emph{horizon crossing}. The tensor-to-scalar ratio ($r$) and scalar tilt ($n_{s}$) can be computed as
\begin{equation}
r=16\epsilon\left(\varphi_{hc}\right)\,,
\label{eq:r}
\end{equation}
\begin{equation}
n_{s}=1-6\epsilon\left(\varphi_{hc}\right)+2\eta\left(\varphi_{hc}\right)\,.
\label{eq:ns}
\end{equation}
Assuming $50<N_{*}<60$, we get $5.96\times10^{-6}<r<1.02\times10^{-5}$ and $0.941<n_{s}<0.951$ for the $\gamma=2/3$ case. According to Planck 2018 results, $r<0.064$ and $0.957<n_{s}<0.968$.

\section{6. Concluding remarks}
I have presented a novel approach to describe early inflation and the late time acceleration of the universe in a unified picture, in the sense that a single scalar potential can account for both cosmic eras. In this proposal, the potential is determined by imposing that a scalar field with a particular meaning is an exact solution to the Friedmann and Klein-Gordon equations in a spatially flat FLRW universe. \\\\
The ansatz for the scalar field has been associated with the energy of the spacetime foam inside the Hubble horizon. This correspondence leads to a potential that allows for more than enough inflation in the early universe, and it yields acceptable values for the inflationary observables. Demanding consistency with the observational data and imposing natural bounds on the parameters of the model, one gets the following estimate for the mass of the inflaton: $m\sim10^{-5}M_{p}$. Moreover, the potential can also describe the current expansion of the universe as the field ends up at a circle of vacua at positive energy density given by (\ref{eq:vacuum energy}), thus acting effectively as a cosmological constant. Although the model cannot justify the exact value of the vacuum energy, it predicts that it has to be much smaller than $M_{p}^{4}$ if one assumes that the Hubble radius during inflation is of the order of the Compton wavelength of the inflaton. \\\\
Remarkably, the requirement that the potential does not blow up for finite values of the field yields the quantization of $\gamma$. The particular solution $\gamma=2/3$ turns out to be special, because it is the one that minimizes the mass of the inflaton. Since this is precisely the solution with direct connection with the holographic model of spacetime foam, this might be regarded as an argument to strengthen our faith on the K\'arolyh\'azy uncertainty relation.      

\section{Acknowledgements}
I would like to express my sincere gratitude to Y. Jack Ng and Diego Pav\'on for their support and insightful words. I am also grateful to L.N. Granda, Thomas W. Kephart and Ahmad Sheykhi for useful comments.

\bibliographystyle{apsrev4-1} 
\bibliography{model_unified_inflation_and_dark_energy_october2022.bib} 

\end{document}